\documentclass[11pt]{article}
\usepackage{url}
\usepackage[sectionbib]{natbib}
\usepackage{graphicx}
\usepackage{amssymb}
\usepackage[figuresright]{rotating}
\usepackage{txfonts}
\usepackage{cospar}
\renewcommand{\bibsection}{\section*{REFERENCES}}
\renewcommand{\bibhang}{\setlength{8mm}}
\renewcommand{\bibsep}{\setlength{0mm}}
\def\ros{{\sl ROSAT }}
\def\cha{{\sl Chandra }}
\def\ergsec{\hbox{erg s$^{-1}$ }}
\def\ergcm{\hbox{erg cm$^{-2}$ s$^{-1}$ }}

\def\am{$^{\prime}$}
\def\as{$^{\prime\prime}$}
\def\psr{PSR 0628-28 }
\def\xr#1{\parindent=0.0cm\hangindent=0.8cm\hangafter=1\indent#1\par}

\title{{\em{Chandra}} OBSERVATIONS OF TWO PULSARS PSR 0628-28 AND PSR 1813-36}

\author{Hakk{\i} \"{O}gelman\address{Department of Physics, 
University of Wisconsin-Madison, 1150 University Avenue, Madison, 
WI 53706}, Emre Tepedelenlio\v{g}lu$^{1}$}

\begin{document}

\maketitle

\begin{abstract} 
\psr is an X-ray emitting
radio pulsar which was observed with Advanced CCD Imaging Spectrometer
(ACIS) on board \cha on 2001 November 04 and on 2002 March 25 for 2000 s
and 17000 s, respectively. The source countrate was 0.0111$ \pm$0.0013cps. 
Making \psr to be the longest period X-ray emitting pulsar. 
The spectral distribution of counts can be described by several model fits. 
A blackbody fit yields a temperature $kT=2.69_{-0.09}^{+0.09}\times 10^{6}$K 
together with $N_{H}=0.082_{-0.03}^{+0.04}\times 10^{22}$cm$^{-2}$ and 
a powerlaw fit yields a photon index of $\gamma =2.65_{-0.29}^{+0.28}$ 
with a hydrogen column density of $N_{H}=0.13_{-0.021}^{+0.028}\times 10^{22}$cm$^{-2}$. 
Confirming the previous \ros pointed observation for PSR 1813-36, 
there was no positive detection from the 30ks \cha ACIS observation on 2001 
October 25. We obtained an upper limit for the countrate of $2.3\times 10^{-4}$cps.
\end{abstract}

\section{INTRODUCTION}

Radio channel is the best way to discover neutron stars, but their 
characteristics are best revealed via the X-ray observations i.e. 
whether it's a cooling neutron star or not.
With this in mind, we observed with \cha two \ros All-Sky-Survey (RASS) 
pulsars, that were claimed to be detected, to confirm their detection 
and examine their X-ray pulses and spectra.
During RASS PSR 1813-36 and \psr had detection likelihood just below the 
detection threshold, Becker et al. (1992). Therefore both 
pulsars had follow-up \ros pointed observations. PSR 1813-36 was not 
detected and we also report a non-detection of the source for the \cha observation. 
However \psr was detected with $\sim$60 counts in the \ros pointed observation and here we 
report a positive detection of a point source around the pulsar's radio position. 
Due to low countrate spectral analysis did not suffice to distinguish between 
various pulsar models such as, powerlaw spectrum or a blackbody spectrum.
In this article we report on observations acquired for both pulsars, namely 
PSR 1813-36 and \psr. 
\section{OBSERVATIONS}
\subsection{\psr}
\psr was observed twice, one on 2001 November 04 and the other on 2002 
March 25 for 2000 s and 17000 s, respectively. For both observations photons were 
collected using the Advanced CCD Imaging Spectrometer (ACIS; Burke et al. (1997)), which has 
resolution $\Delta E/E\sim 0.1$ at 1 keV scaling as 1/$\sqrt{E}$ over its 
0.2-10 keV range. The pulsar was positioned on the back-illuminated S3 chip 
of the ACIS-S array, offset by 0\am.58 from the aim point, where the 
PSF is undersampled by the 0\as.4920$\times$0\as.4920 CCD pixels. 
Data were collected in the nominal timing mode, with 1.141s exposures 
between CCD readouts, and in "FAINT" spectral mode. 
Data reduction and analysis were performed by CIAO data analysis software 
package, version 2.2.1. We reprocessed the Level 1 event 
data to correct the detrimental effects of charge transfer efficiency. 
The imaging, timing and spectral analysis presented here were done 
only on the 17 ks observation, the 2.0 ks observation was disregarded due 
to an interruption by a large solar storm. The background countrate during this
solar storm increased by a factor of $\sim$20. The net source countrate during the first 
observation period was $0.0159 \pm 0.0050$ where the error is in the 90\% confidence
range. Since by taking into account the 2.0 ks observation 
we gain only $\sim$27 counts which is not enough to improve our statistics significantly 
we prefered to disregard these counts, which potentially can be misleading.

The net countrate of the 17 ks observation is $0.0111 \pm 0.0013$. 
We also analysed the \ros archival data and determined the 
\ros countrate of \psr which after normalizing to \cha countrate is 
$0.0120\pm 0.0025$. The countrate of the portion of the observation that we used
as can be seen is consistent within the errors with both 2.0 ks and \ros observations.
The measured PSF of \psr is consistent with the ACIS point-source 
response, hence the ACIS image reveals a pointlike X-ray source at the 
pulsar position. The Chandra position of \psr is 
$\alpha=06^{h}30^{m}49^{\prime\prime}.427$, 
$\delta=-28^{\circ}34^{\prime}43^{\prime\prime}.60$ (J2000.0), which 
considering 0.\as5 rms error and $\sim$0.\as6 ~absolute astrometric 
accuracy of \cha, is in good agreement with the radio position of 
$\alpha$=06$^{h}$30$^{m}$49{\as}.531, 
$\delta=-28^{\circ}34^{\prime}43^{\prime\prime}.60$ (J2000.0),
Taylor et al. (1993). We should also note that the \cha position is also consistent with the 
\ros position of the \psr. We also took into account the proper motion of the
pulsar. The \cha position of the pulsar is $\sim$0.\as5 away from the expected 
position at the observation epoch which is extrapolated from the proper 
motion measurements taken from  Taylor et al. (1993). 

With this positional coincidence we can firmly say that \psr is one of the handful 
of X-ray emitting radio-pulsars.
\subsection{PSR 1813-36}
PSR 1813-36 was observed on 2001 October 25 for 30 ks with ACIS-S. 
Confirming the previous result for the \ros pointing observation no 
source was detected at the pulsar posistion of 
$\alpha$=18$^{h}$17$^{m}$05{\as}.76, 
$\delta$=-36$^{\circ}$18{\am}05{\as}.50 (J2000.0), Taylor et al. (1993). 
The observation yielded a background countrate per pixel square of  
3.4$\times$10$^{-6}$ cps/pixel$^{2}$, hence for a 50 pixel$^{2}$ extraction 
region and a signal to noise ratio of 3 the source should have minimum 
countrate of 2.3$\times$10$^{-4}$ cps for detectability. Assuming a 
powerlaw like spectrum this countrate gives an upperlimit for the flux of 
1.5$\times$10$^{-15}$ \ergcm, which is in the sensitivity range of ACIS-S 
  (0.2-10 keV), for a detector ontime of 30 ks. In 
the same range the implied X-ray flux from spin-down energy is 
2.1$\times$10$^{-15}$ \ergsec, which corresponds to a luminosity of 
$L_{X}$=3.7$\times$10$^{30}$ \ergsec at an assumed distance of 3.8 kpc. 
\section{RESULTS FOR \psr}
\subsection{Timing}
To search for pulsations from \psr 184 source counts were extracted from a
2\as ~radius aperture around the \cha position of the pulsar. The photon 
arrival times were barycenter corrected using CIAO axbary program. Then the 
$Z^{2}_{1}$ statistic (Buccheri et al. (1983)) was calculated over a 
narrow range of frequencies  at the expected radio frequency 
($f=0.80358910$ Hz, Taylor et al. (1993)) The resulting periodogram peaked at a
value of $Z^{2}_{1}=11.8961$ at $f=0.803352585\pm 0.000059$ Hz referenced
to the observation epoch MJD 52,358.9. Which is not consistent with the
value obtained for the frequency from extrapolation of the radio ephemeris 
($f$,$\dot{f}$). A wider search including the ACIS sampling frequency of
$f$=0.8763934 Hz, showed a number of high peaks, indicating that these
results are influenced by the beat frequencies between the pulsars
rotation  and the \cha ACIS  sampling. Continuing to investigate
the folding results only around the expected pulsar frequency,
 we get $Z_{1}^{2}$=0.7 which is
distrubuted like $\chi^{2}$ with 2 dof, hence the probability of getting
this value or higher is 0.713 which is not too remarkable. Looking at
the higher harmonics, $Z_{2}^{2}$=0.92 (the sum of 1st \& 2nd
harmonics), which has a random probability of 0.92 and $Z_{3}^{2}$=10.1
which has a random probability of 0.12. The conclusion which can be 
drawn from this 
is that the data was not sufficient to claim the existence of pulsed-emission.
\subsection{Spectral}
To fit a spectral model to the pulsar data we extracted the source counts 
from a circle of radius 2\as ~and extracted the background photons from a 
box of dimensions 131\as$\times$135\as, which was taken considerably away 
from the pulsar where no other sources were present. The extraction radius 
of 2\as ~contains 90\% of the events below 2 keV. 
The countrates for the source and background were 0.0111 cps and 0.214 cps, 
respectively. 
The CIAO tool MKARF and MKRMF were used to build the corresponding 
point-source mirror response matrix. The extracted photons then were 
binned with 0.5 keV intervals. The resulting spectra were 
fitted using the generalized fitting engine of CIAO, {\em Sherpa}. Two 
different models were tried, a blackbody and a powerlaw model. 
For the blackbody fit the $N_{H}$ vs. $kT$ error ellipses have been 
plotted in Figure 1.
\begin{figure}[t]
\begin{center}
\includegraphics[width=95mm]{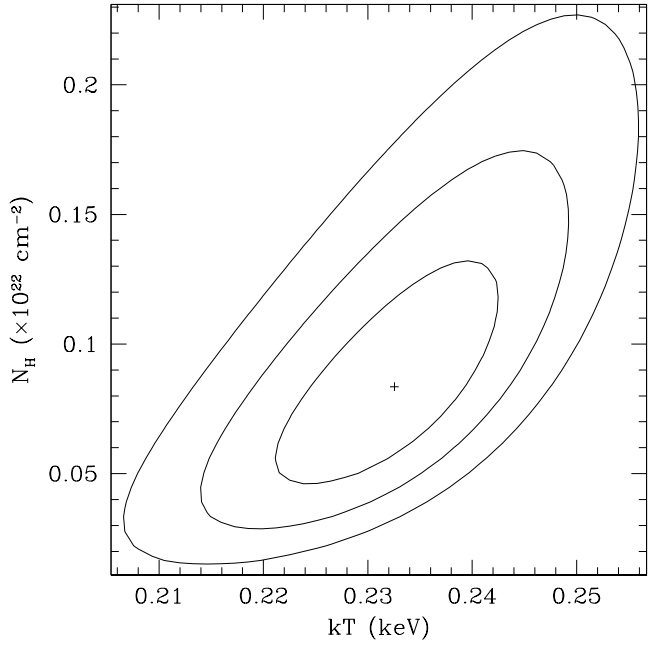}
\end{center}
\caption{Error ellipses of $N_{H}$ vs. $kT$, the contours are for 
1, 2 and 3$\sigma$ uncertainities}
\end{figure}%
The hydrogen column density that we obtained 
from the dispersion measure is $N_{H}=1.07\times 10^{21}$ cm$^{-2}$, where
we assumed a ratio of $n_{H}/n_{e}=10$. The blackbody fit had a 
reduced $\chi^{2}$ of 0.29. The best
fit parameter for this fit was $kT=0.232$ keV for the
surface temperature of  the neutron star and $N_{H}=8\times 10^{20}$ 
cm$^{-2}$ for the hydrogen column density (see Table 1).  
This model gives an absorbed flux of 2.04$\times$10$^{-14}$ \ergcm 
in the 0.1-10 keV range, with a distance of 2.14 kpc this flux 
corresponds to a X-ray luminosity of $L_{X}$=1.13$\times$10$^{31}$ 
\ergsec, in the same range. The implied source radius is very small (0.17 km); possibly from polar caps.

 A powerlaw model also gave an  acceptable fit. 
The reduced $\chi^{2}$ minimized at a value of 0.32 for 
the parameters $\gamma=2.65$ and $N_{H}=1.29\times 10^{21}$ cm$^{-2}$.
This fit yielded a flux of 4.06$\times$10$^{-14}$ \ergcm, 
which with a distance of 2.14 kpc corresponds 
to a X-ray luminosity of $L_{X}$=2.25$\times$10$^{31}$ \ergsec (see Figure 2).

\begin{table}[b]
\begin{center}
\caption{\vfill
Model parameters for \psr. The errors are the 90\% confidence level range.}
\begin{tabular*}{1.0\textwidth}{@{\extracolsep{\fill}}lccr}\hline
Model & &  Flux in \ergcm & $N_{H}$ in cm$^{-2}$  \\
\hline
\\
Blackbody & $kT=0.232^{+0.008}_{-0.008}$ keV & $2.04^{-0.29}_{+0.29}\times$10$^{-14}$ & $0.08^{-0.03}_{+0.04}\times 10^{22}$ \\
\\
Powerlaw & $\gamma=2.65^{+0.28}_{-0.28}$ & $4.06^{-0.59}_{+0.59}\times$10$^{-14}$ & $0.13^{-0.02}_{+0.03}\times 10^{22}$ \\
\\
\hline
\end{tabular*}
\end{center}
\end{table}

\begin{figure}
\begin{center}
\includegraphics[width=95mm]{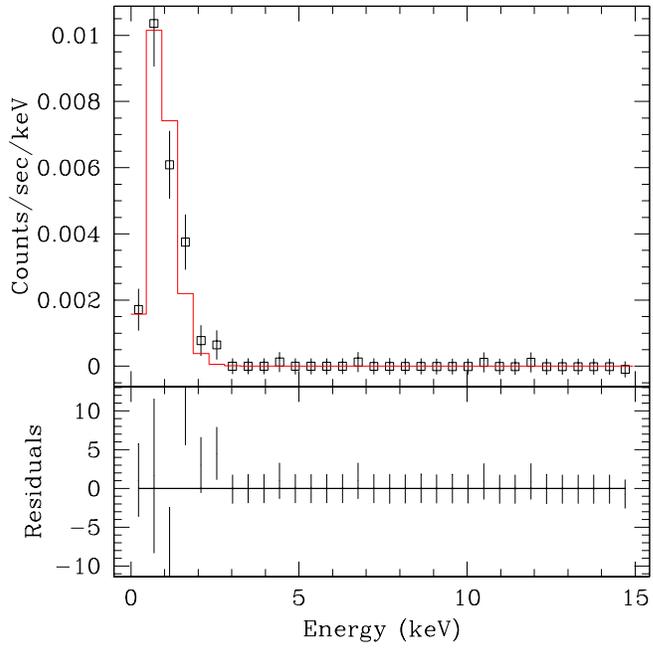} 
\includegraphics[width=95mm]{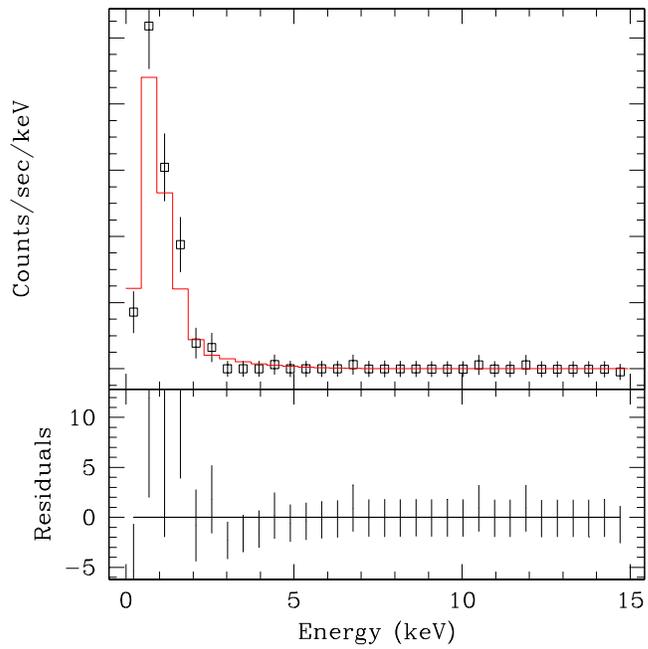}\\
\end{center}
\caption{Spectrum of \psr along with ({\em top)} 
blackbody  fit and  ({\em bottom}) powerlaw fit. The bottom panel for both 
are the counting residuals.}
\end{figure}
\begin{figure}
\begin{center}
\includegraphics[width=90mm,angle=270]{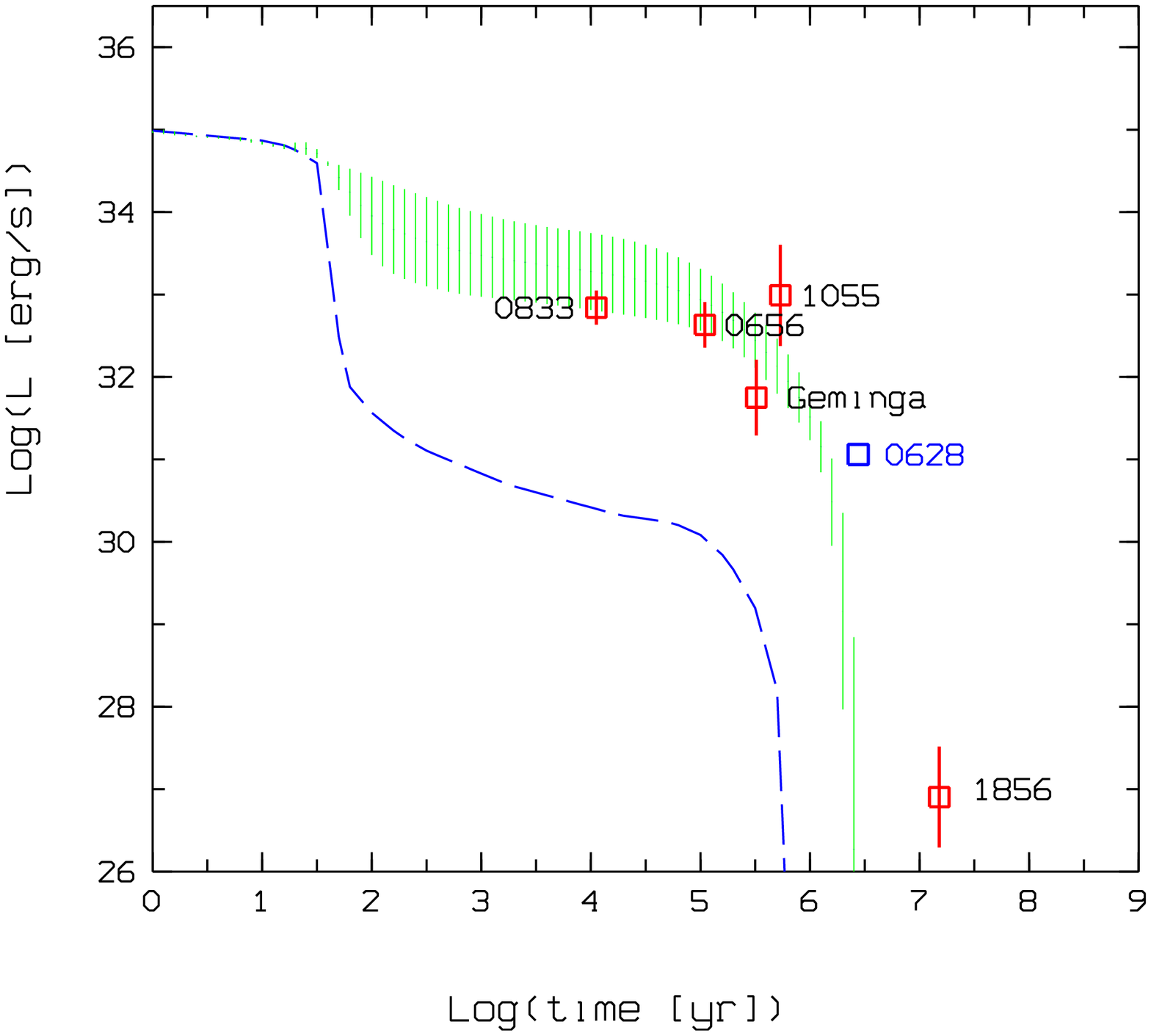}
\end{center}
\caption{The cooling curve for neutron stars and \psr,
where it's luminosity is derived from the blackbody fit. The dashed line 
is the exotic cooling curve and the shaded region is various standard
cooling curves.}
\end{figure}

\section{SUMMARY \& DISCUSSION} 
We report the discovery of X-ray emission from a long-period, 
(1.24s) radio-pulsar with a mild rotational energy loss-rate. Expressed in 
numbers, the measured power-law fit $L_{X}$=2.25$\times$10$^{31}$ \ergsec 
which should be compared to the 0.1\% of $\dot{E}$, a robust formula which 
gives a good fit to the rotation-powered pulsars (Becker \& Tr\"{u}mper (1997)). The 0.1\%
of $\dot{E}$ is $10^{29}$\ergsec. Effectively the observed X-ray
Luminosity is 100 times more than that predicted by the magnetospheric
model. We should also note that the blackbody fit luminosity is about the
same as the powerlaw luminosity. What is the  source of this extra
luminosity? \psr is an extreme pulsar long-period, small $\dot{E}$. The
physical process producing the X-rays may include  reheating due to vortex
creep, accretion from a fall-back disc, magnetic decay like the AXPs or
accretion from the ISM. Also the luminosity derived from the blackbody fit
($L_{X}$=1.13$\times$10$^{31}$\ergsec) and the characteristic age
($P/2\dot{P}$=2.8$\times$10$^{6}$ yr) of the pulsar puts it on the
standard cooling curve, see Figure 3, although it is on the fast-falling,
photon-cooling dominated part.

Deeper observations, to produce sufficient counts with better timing
accuracy, are needed in order to resolve the issues stated above. \cha
observations have focused our attention on this extreme pulsar.

\section{ACKNOWLEDGEMENTS}
We thank Patrick O. Slane, Werner Becker, Roberto Mignani for useful 
discussions. This work has been supported by NASA Grant \#02500773.

\bibliography{bibfile}
\section{REFERENCES}

\xr{Becker W., J. Tr\"{u}mper, The X-ray luminosity of rotation-powered neutron stars, {\it Astronomy \& 
Astrophysics} {\bf 326}, 682-691, 1997.} 
\xr{Becker W., J. Tr\"{u}mper, H. \"{O}gelman, Search for cooling neutron stars in the \ros survey, in 
{\it Isolated Pulsars}, edited by Van Riper K. A., Epstein R. and Ho C., 104, Cambridge 
University Press,  1992.}
\xr{Buccheri R., K. Bennett , G. F. Bignami et al., Search for pulsed $\gamma$-ray emission from the radio 
pulsars in the COS-B data, {\it Astronomy \& Astrophysics} {\bf 128}, 245-251, 1983.}
\xr{Burke, B. E., J. Gregory, M. W. Bautz et al., {\it IEEE  Trans. Electron Devices}, {\bf 44}, 1633, 1997.}
\xr{Taylor J. H., R. N. Manchester, A. G. Lyne, Catalog of 558 pulsars, {\it Astrop. J. Suppl. Ser.} 
{\bf 88}, 529-568, 1993.}

\end{document}